\documentclass[a4paper]{article}
\usepackage{amsmath}
\usepackage{latexsym}
\usepackage{graphicx}
\usepackage{cite}
\begin{document}

\title{\textbf{ \Large{Note on Nariai and Tomita's and Starobinsky's 
cosmological solutions in the $R^2$ modified gravity}}}

\author{Kenji Tomita \\
{\small Yukawa Institute for Theoretical Physics, Kyoto University, Kyoto 606-8502, 
Japan} \\ {\small Email: ketomita@ybb.ne.jp}}
\maketitle

\begin{abstract}
Cosmological solutions derived by Nariai and Tomita (1971) and by Starobinsky (1980)
are compared, and it is shown that the former derived de Sitter expansion 
in the $R^2$ modified gravity (without cosmological constant) at the earliest stage,
and nine years later the latter derived the well-known inflationary solution.
Next the property of their simplified models is described using the method of conformal 
transformations, and how the inflation arises and the singularity is avoided is shown. Finally the 
initial and final states of the inflation are discussed. 
\end{abstract}

\section{Introduction}
In 1971, Nariai[1] formulated gravitational equations with Lagrangian density
$L_g$ including the square of Ricci tensors,  in terms of an extended version of the 
renormalized  theory of gravitation due to Utiyama and DeWitt.[2]
He aimed to derive a regular solution without any
 cosmological singularity, and Nariai and Tomita[3] showed many examples of the regular
  solutions, solving the equations in the homogeneous and isotropic model. They found that,
as the time increases, most of them do not tend to the Friedmann model, but to the de Sitter 
model in the future.
This may be the first time when we found that the square of Ricci tensors in the Lagrangian 
causes de Sitter expansion without cosmological constant.

In 1980, Starobinsky[4] formulated independently gravitational equations with $L_g$
 including  the square of Ricci tensors, and derived the well-known solution  in which the universe 
 starts with the de Sitter expansion, and the expansion rate decreases with time. This behavior 
 of the universe was taken notice a few years later as one of the models representing the
 inflation of the universe.[5-9]
 
After several years, a new system of simplified gravitational equations with $L_g = R + 
\alpha R^2$ \ was considered[10, 11], where $R$ is the Ricci scalar and $\alpha$
 is a constant, and next to solve them, a method using conformal transformations was 
 exploited[12-15] and the behavior of the models was shown more clearly.

In this Note, I first compare the equations to be solved for deriving the models which were
given by Nariai and Tomita  and by Starobinsky, and discuss the difference in their solutions. 
Next I describe their  simplified models, and use conformal transformations to find the 
behaviors  of the models and show how the inflation is caused and the singularity is avoided.
 Finally, I discuss the initial and final states of the inflation.

\section{Comparison of Nariai and Tomita's and Starobinsky's treatments}
In Nariai and Tomita's treatment[3], the gravitational Lagrangian density is
\begin{equation}
  \label{eq:a1}
L_g = \sqrt{-g} [R + \eta (R^2 + \tilde{\alpha} R_{\mu\nu} R^{\mu\nu})],
\end{equation}
where $R_{\mu\nu} $ and $R$ are the Ricci tensor and Ricci scalar, and $\eta$ and $\tilde{\alpha}$
are constants. Variating the action integral $I = \int (\kappa^{-1} L_g + L_m) d^4 x$ with respect to
 $g^{\mu\nu}$, we obtain the gravitational equations, where $\kappa$ is Einstein's gravitational
constant  and $L_m$ is the matter Lagrangian density.

In the spatially flat cosmological space-time with the line-element
\begin{equation}
  \label{eq:a2}
ds^2 = c^2 dt^2 - a^2(t) (dr^2+r^2 d\theta^2 +r^2 \sin^2 \theta d \phi^2),
\end{equation}
we obtain 
\begin{equation}
  \label{eq:a3}
 \begin{split} 
8\pi G\rho = & 3(\ddot{a}/a)^2 + 3 (t_c)^2 [\{ \ddot{a}/a + (\dot{a}/a)^2\} 
\{ \ddot{a}/a - (\dot{a}/a)^2\} \\
&  -2  (\dot{a}/a) \{\ddot{a}/a + (\dot{a}/a)^2\} ^.] ,
\end{split}
\end{equation}
\begin{equation}
  \label{eq:a4}
\dot{\rho} + 3(\dot{a}/a) (\rho +p)^2 = 0,
\end{equation}
where a dot denotes $d/dt$,  $\rho$ and $p$ are the density and pressure of perfect fluid,
$G$ is the Newtonian gravitational constant, and $c = 1$.  The constant $t_c$ is defined by
\begin{equation}
  \label{eq:a5}
(t_c)^2 \equiv -2 (\tilde{\alpha} + 3) \eta.
\end{equation}
Here it is not evident whether $(t_c)^2$ is positive or negative. In the radiation-dominated case
 ($p = \rho/3$), we obtain from Eqs.  (3) and (4) 
\begin{equation}
  \label{eq:a6}
 \begin{split} 
(\dot{a}/a)^2 &+ (t_c)^2 [\{ \ddot{a}/a - (\dot{a}/a)^2\}^2 
- 2 (\dot{a}/a) \{ {\ddot{a}}^./a - (\dot{a}/a)^3\}] \\
&  = (8\pi G\rho_c/3) (a_c/a)^4,
\end{split}
\end{equation}
which is rewritten as
\begin{equation}
  \label{eq:a61}
 \begin{split} 
(\dot{a}/a)^2 &= (t_c)^2 [2\dot{a} {\ddot{a}}^./a^2 - (\ddot{a}/a)^2 + 2\ddot{a} {\dot{a}}^2/a^3
- 3(\dot{a}/a)^4] \\
&  + (8\pi G\rho_c/3) (a_c/a)^4,
\end{split}
\end{equation}
where $\rho a^4 = \rho_c {a_c}^4$ and $\rho_c$ and $a_c$ are constants.
Solving this equation in the future direction and the past direction, Nariai and Tomita[3] found 
that, in the case of $(t_c)^2 > 0$, we can obtain regular solutions in which the universe
bounces at the maximum density, but most solutions tend to the de Sitter solution with 
 exponential expansion as $t$ increases. In the case of $(t_c)^2 < 0$, on the other hand, all
  solutions show exponential expansion in the past and cannot bounce. So they did
  not analyze the solutions in this case more, because they paid attention only to the
  bouncing of the universe.  However they found probably first that in spite of the signature of $(t_c)^2$
the exponential expansion  is caused by the square of the Ricci tensor without cosmological
 constant. 
  
In Starobinsky's treatment[4], the gravitational Lagrangian density is similar 
to that of NT. In the cosmological space-time with the spatial curvature 
$K = 1, 0,-1$,  and without matter, he obtained  
\begin{equation}
  \label{eq:a7}
 \begin{split} 
(\dot{a}^2+K)/a^2 &= \frac{1}{H^2} [(\dot{a}^2+K)/a]^2 \\
&-\frac{1}{M^2}
[2\dot{a} {\ddot{a}}^./a^2 - (\ddot{a}/a)^2 + 2\ddot{a} {\dot{a}}^2/a^3 - 3(\dot{a}/a)^4
-2K {\dot{a}}^2/a^4 + K^2/a^4],
\end{split}
\end{equation}
where $H$ and $M$ are constants (similar to $\eta$ and $\tilde{\alpha}$) appearing in
the gravitational Lagrangian density. Comparing Eqs. (7) and  (8),
we find that, if $K = 0$, the two treatments have the same main terms, if we replace $1/M^2$ by
$- (t_c)^2$. The another term with $1/H^2$ does not appear in the treatment of NT 
and seems to have a role to strengthen the de Sitter expansion by assuming $H^2 > 0$.
In the case of $M^2 > 0$ (i.e. $(t_c)^2 < 0)$, Starobinsky finds the solution in which
the universe starts from the de Sitter expansion and its expansion slows down with time. 
In the case of $M^2 < 0$ (i.e. $(t_c)^2 > 0)$, the de Sitter expansion grows with time or
is ``future stable''. It is found therefore that the behaviors in both treatments are consistent,
though they may have been interested in different aspects.

\section{Simplified treatment and the conformal transformation}
We start with the simplified gravitational Lagrangian density
\begin{equation}
  \label{eq:b1}
L_g = \sqrt{-g} f(R)
\end{equation}
with
\begin{equation}
  \label{eq:b2}
f(R)  = R + \alpha R^2,
\end{equation}
where $\alpha$ is a constant. This $f(R)$ is given in a special case of $\eta = \alpha$ and 
$\tilde{\alpha} = 0$ in the Lagrangian density of NT.
The field equation is derived by variating the action integral $I = \frac{1}{2\kappa^2} \int L_g  d^4 x$
\ (without matter) with respect to $g_{\mu\nu}$, we obtain
\begin{equation}
  \label{eq:b3}
F(R) R_{\mu\nu} - \frac{1}{2} f(R) g_{\mu\nu} -\nabla_\mu \nabla_\nu F(R)
+ g_{\mu\nu} \Box F(R) = 0, 
\end{equation}
where I used De Felice and Tsujikawa's notation[15] as $F(R) \equiv df/dR = 
1 + 2\alpha R$, \ $\nabla_\mu$ is the covariant derivative,  \
$\Box F = (1/\sqrt{-g})  (\sqrt{-g} g^{\mu\nu} F_{,\nu})_{,\mu}$ and $,\mu$ denotes 
$\partial/\partial x^\mu$. Rewriting Eq.(11), we obtain
\begin{equation}
  \label{eq:b4}
G_{\mu\nu} \equiv R_{\mu\nu} - (1/2) g_{\mu\nu} R = \kappa^2 {T^{(G)}}_{\mu\nu},
\end{equation}
\begin{equation}
  \label{eq:b5}
 \begin{split} 
\kappa^2 {T^{(G)}}_{\mu\nu} & \equiv   \frac{1}{2}g_{\mu\nu} (f - R) + \nabla_\mu \nabla_\nu F
 -g_{\mu\nu} \Box F + (1 - F) R_{\mu\nu} \\
& = \alpha [\frac{1}{2} g_{\mu\nu} R^2 + 2\nabla_\mu \nabla_\nu R - 2g_{\mu\nu} \Box R -2 R R_{\mu\nu}].
\end{split}
\end{equation}
Thus $\alpha R^2$ in $f(R)$ behaves as a kind of matter fields.

Next, let us transform conformally as
\begin{equation}
  \label{eq:b6}
\tilde{g}_{\mu\nu} = F g_{\mu\nu}
\end{equation}
and introduce a scalar field $\phi$ as
\begin{equation}
  \label{eq:b7}
\kappa \phi \equiv \sqrt{3/2} \ln F = \sqrt{3/2} \ln (1+ 2\alpha R).
\end{equation}
Then the action in the Einstein frame reduces to
\begin{equation}
  \label{eq:b8}
S_E = \int d^4 x \sqrt{-\tilde{g}} [\frac{1}{2\kappa^2} \tilde{R} - \frac{1}{2}  \tilde{g}^{\mu\nu}
\phi_{,\mu} \phi_{,\nu} - V(\phi)]
\end{equation}
following De Felice and Tsujikawa[15], where 
\begin{equation}
  \label{eq:b9}
V(\phi) \equiv \frac{FR -f}{2\kappa^2 F^2} + V_0 = \frac{1}{2 \alpha \kappa^2} 
(\frac{\alpha R}{1+2\alpha R})^2 + V_0
\end{equation}
and $V_0$ is a constant. Variating the action (16) with respect to $\phi$, we obtain
\begin{equation}
  \label{eq:b10}
\tilde{\Box} \phi \equiv  (1/\sqrt{-\tilde{g}}) (\sqrt{-\tilde{g}} \tilde{g}^{\mu\nu} 
\phi_{,\nu})_{,\mu} = V_{,\phi}
\end{equation}
which leads to
\begin{equation}
  \label{eq:b11}
d^2 \phi/d \tilde{t}^2 + 3 \tilde{H} d\phi/d\tilde{t} + V_{,\phi} = 0,
\end{equation}
where $d \tilde{t} \equiv \sqrt{F} dt, \tilde{a} \equiv \sqrt{F} a$ and $\tilde{H} \equiv 
d \ln \tilde{a}/{d \tilde{t}}$. Using Eq. (15), $F, R$ and $V$ are expressed in 
terms of $\phi$ as
\begin{equation}
  \label{eq:b12}
F = {\rm e}^{\sqrt{2/3} \kappa \phi}, \qquad \alpha R = \frac{1}{2} ({\rm e}^{\sqrt{2/3} \kappa \phi} -1) ,
\end{equation}
\begin{equation}
  \label{eq:b13}
V = \frac{1}{8 \alpha\kappa^2} ( 1- {\rm e}^{-\sqrt{2/3} \kappa \phi})^2 + V_0.
\end{equation}

\noindent In the case of $\alpha = \alpha_0 (> 0)$, $V$ is expressed as
\begin{equation}
  \label{eq:b14}
V = \frac{1}{8 \alpha_0 \kappa^2} ( 1 - {\rm e}^{-\sqrt{2/3} \kappa \phi})^2 
\end{equation}
by taking $V_0$ so that $V = 0$ for $\phi = 0$.

\noindent In the case of $\alpha = - \alpha_0 (< 0)$, $V$ is 
\begin{equation}
  \label{eq:b15}
V = \frac{1}{8 \alpha_0 \kappa^2} [2 - (1 -{\rm e}^{-\sqrt{2/3} \kappa \phi})^2] 
\end{equation}
by taking $V_0$ so that $V = 1/(8 \alpha_0 \kappa^2)$ for $\phi \rightarrow \infty$.
These behaviors of $V (\phi)$ in the cases of $\alpha = \alpha_0$ and $-\alpha_0$ are
shown in Figs. 1 and 2, respectively.

\begin{figure}[h]
\caption{\label{fig:1} The potential $V(\phi)$ in the case of $\alpha = 
\alpha_0 > 0$. $v \equiv 
 (8\alpha_0\kappa^2) V(\phi)$ and $u \equiv \sqrt{2/3} \kappa \phi$. Inflation is caused in 
 the range $u \gg 1$.} 
\centerline{\includegraphics[width=8cm]{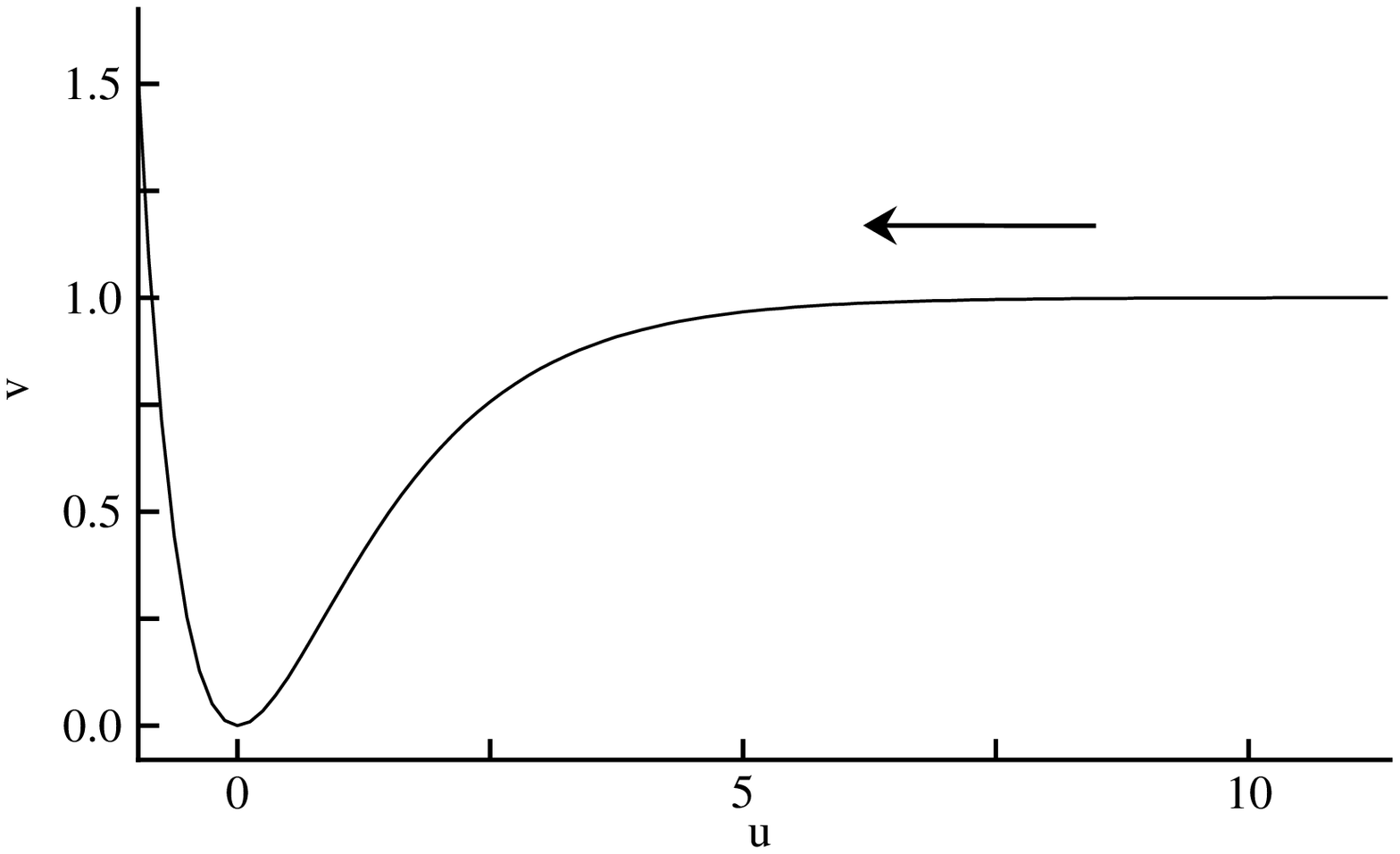}}
\caption{\label{fig:2} The potential $V(\phi)$ in the case of $\alpha = 
- \alpha_0 < 0$. $v \equiv 
 (8\alpha_0\kappa^2) V(\phi)$ and $u \equiv \sqrt{2/3} \kappa \phi$. Inflation is caused in 
 the range $u \gg 1$.} 
\centerline{\includegraphics[width=8cm]{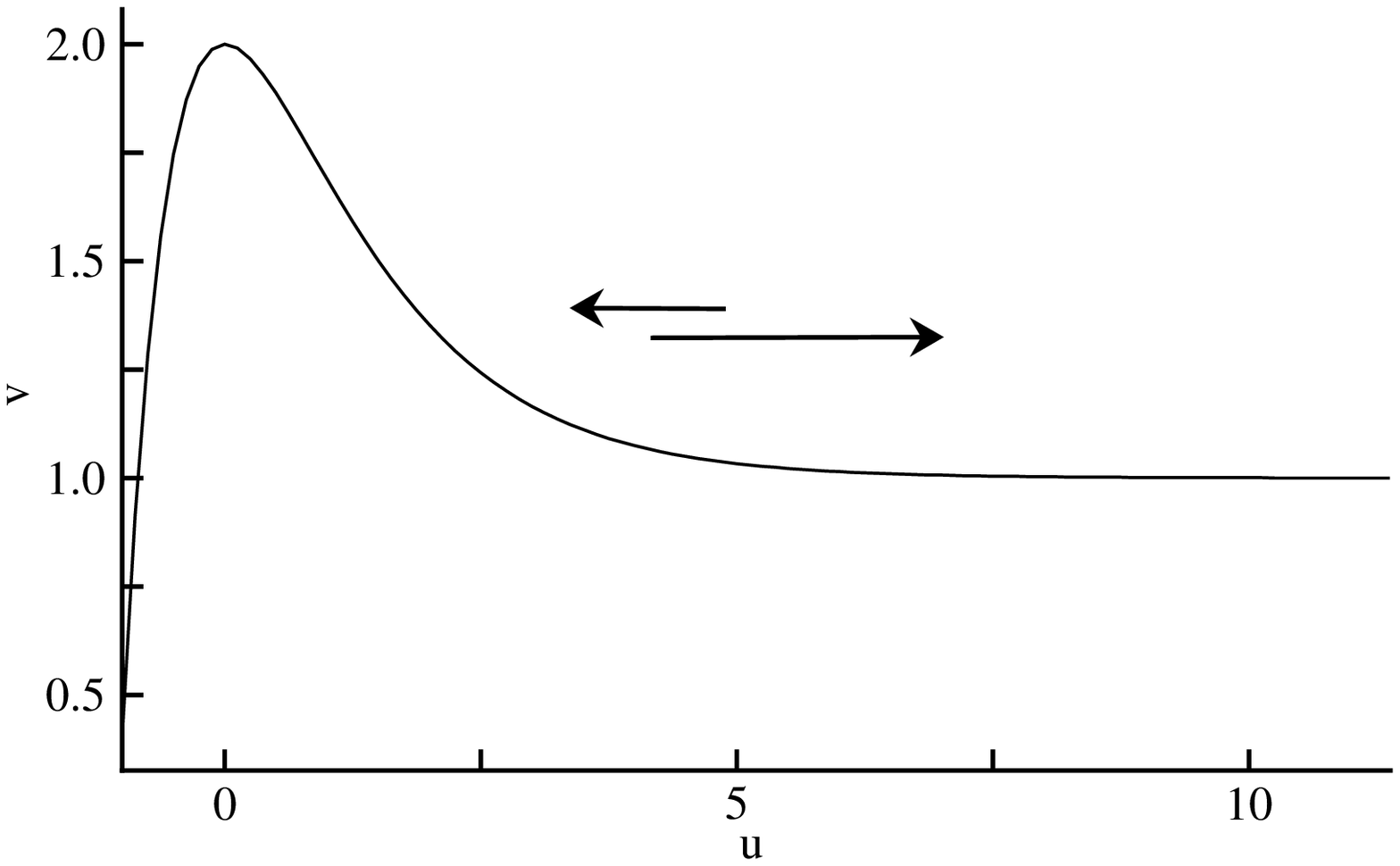}}
\end{figure}

The evolution of the model universes is found as follows through the behavior of $\phi$
 in the potential $V$. At the stage of $\kappa \phi \gg 1$, 
$V$ is nearly constant in both cases of $\alpha = \alpha_0$ and $-\alpha_0$, as shown 
in Figs. 1 and 2, so that the universes show the behavior of slow-roll inflation. 

In the case of $\alpha = \alpha_0$ \ (corresponding to $1/M^2 = -(t_c)^2 > 0$), 
$\phi$ decreases with time after the inflationary expansion,
and, at the stage of $\kappa \phi \simeq 0$, the universe shows the oscillatory behavior and it reaches the Hubble damping.  The inflation in this case is 
eternal in the past direction.

In the case of $\alpha = - \alpha_0$ \ (corresponding to $1/M^2 = -(t_c)^2 < 0$), 
the universe changes around $\kappa \phi \simeq 0$ from
 the contracting stage to the expanding stage by the action of the wall, accelerates the 
expansion as $\kappa \phi$ increases, and reaches the inflationary stage with $\kappa \phi 
\gg 1$ \ (in Fig. 2). The inflation in this case is eternal in the future direction.

\section{Discussions}
In order that the inflations may be compatible with the observational temperature anisotropies
and solve horizon and flatness problems, the number of e-foldings of the inflations must be finite
 ($\sim 70$) [16].  In the case of $\alpha =
  \alpha_0$, we may adjust the number of e-foldings by assuming that the universe started 
from highly anisotropic states and the de Sitter-type inflation began at the epoch when 
anisotropy damped (Maeda[12]).

Now let us consider another possibility of compatibility. So far we have considered separately 
the cases of $\alpha = \alpha_0$ and $-\alpha_0$. If we can
 change the signature of $\alpha$ in $f(R)$ smoothly around an epoch $t_1$ corresponding 
 to $R = R_1$ with
  $\kappa \phi_1 \gg 1$,  the universe expands after the bouncing, inflates with the finite
  e-folding number around the epoch $t_1$ and finally reach the stage of oscillatory damping.
The value of $\phi_1$ may be constrained by the above observational condition on the e-folding.

\end{document}